\documentclass[aps,prc,superscriptaddress,showpacs,amssymb,amsmath,amsfonts,twocolumn,floatfix,nofootinbib]{revtex4-1}
\usepackage{graphicx}

\newcommand{\be}{\begin{eqnarray}}
\newcommand{\ee}{\end{eqnarray}}

\newcommand{\ba}{\begin{array}}
\newcommand{\ea}{\end{array}}

\newcommand{\GWU}{Data Analysis Center at the Institute for Nuclear
    Studies, Department of Physics,
        The George Washington University, Washington, D.C. 20052}
\begin{document}

\title{Baryon photo-decay amplitudes at the pole}

\author{R.~L.~Workman$\,^a$, L. Tiator$\,^b$ and A. Sarantsev$\,^c$}
\affiliation{ $^a$\GWU\\
$^b$Institut f\"ur Kernphysik, Johannes-Gutenberg Universit\"at Mainz, Germany\\
$^c$Helmholtz-Institut, Universit\"at Bonn, Germany
}

\vspace{5cm}
\date{\today}

\begin{abstract}
We derive relations for baryon photo-decay amplitudes both for the
Breit-Wigner and the pole positions. With an updated SAID partial
wave analysis, technically similar to the earliest Virginia Tech
analysis of photoproduction data, we compare photo-decay amplitudes
at both resonance positions for a few selected nucleon resonances.
Comparisons are made and a qualitative similarity, seen between the
pole and Breit-Wigner values extracted by the Bonn-Gatchina group,
is confirmed in the present study.
\end{abstract}

\maketitle

\section{Introduction}
\label{sec:intro}

Baryon resonance properties, evaluated at the pole position, are
beginning to supersede and replace quantities which have generally
been determined using Breit-Wigner (BW) plus background
parameterizations. This is reflected in the most recent~\cite{pdg}
Review of Particle Properties, with many pole values coming from the
recent Bonn-Gatchina multi-channel analyses~\cite{bnga}. While the
pole extraction is well-defined and less model-dependent than the
Breit-Wigner approach, the continuation of fit amplitudes to the
pole is itself a possible source of error. This has motivated
numerous studies involving speed plots, Laurent series
representations, regularization methods, and contour
integration~\cite{ceci11,masj11,yang11,tiat10,suzu10,svar12}. Here
we will compare Breit-Wigner and pole extractions, using an early
SAID fit form, with the focus on N* photo-decay amplitudes.

As the amplitude itself becomes infinite at the pole, we are
interested in residues. We first clarify the connection between
multipole residues and the photo-decay amplitudes.
This can be related to a result published
with the first SAID photoproduction fits~\cite{vpi90}. A comparison
of recent results and the first attempts tabulated in
Ref.~\cite{vpi90} reveal some large discrepancies. We study and
resolve this problem below.

The photo-decay amplitudes determined via Breit-Wigner and pole
methods, as given by the Bonn-Gatchina group~\cite{bnga}, tend to be
very similar in modulus. We reproduce this trend within the original
SAID photoproduction model.

\section{Breit-Wigner versus pole quantities}
\label{sec:bwvpoles}

The total cross section of pion photoproduction can be written in
terms of helicity multipoles by
\begin{eqnarray}
\sigma_{\gamma,\pi}&=&\frac{1}{2}
(\sigma_{\gamma,\pi}^{1/2}+\sigma_{\gamma,\pi}^{3/2})\,,\\
\sigma_{\gamma,\pi}^{h}&=&4\pi\frac{q}{k}\sum_{\alpha(\ell,J,I)}
(2J+1)\,|\mathcal{A}_\alpha^{h}|^2\;C^2\,,\label{sigtot}
\end{eqnarray}
with $q$ and $k$ being the center-of-mass pion and photon momenta. The
factor $C$ is $\sqrt{2/3}$ for isospin $3/2$ and $-\sqrt{3}$
for isospin $1/2$.  The
helicity multipoles are given in terms of
electric and magnetic multipoles
\begin{eqnarray}
\mathcal{A}_{\ell +}^{1/2} & = & -{1\over 2} \left[ (\ell +2)
{E}_{\ell +}
+ \ell {M}_{\ell +} \right] ,\label{helimult1}
\\
\mathcal{A}_{\ell +}^{3/2} & = & {1\over 2} \sqrt{\ell ( \ell + 2)}
\left[ {E}_{\ell +} -{M}_{\ell +} \right] , \label{helimult2}
\\
\mathcal{A}_{(\ell +1)-}^{1/2} & = & -{1\over 2} \left[\ell
{E}_{(\ell +1)-} -
(\ell +2 ) {M}_{(\ell +1)-} \right] , \label{helimult3}
\\
\mathcal{A}_{(\ell +1)-}^{3/2} & = & -{1\over 2} \sqrt{\ell ( \ell
+2)} \left[ {E}_{(\ell +1)-} + {M}_{(\ell +1) -} \right] ,
\label{helimult4}
\end{eqnarray}
with $J=\ell+1/2$ for '$+$' multipoles and $J=(\ell+1)-1/2$ for
'$-$' multipoles, all having the same total spin $J$.

Comparing with the definition of the cross section from a unitary
amplitude~\cite{pdg_kin}
\begin{equation}
\sigma_{i,f}=\frac{4\pi}{k^2}\frac{2J+1}{(2s_1+1)(2s_2+1)}\,|T_{i,f}|^2,
\end{equation}
where $k$ is the c.m. momentum in the initial state and $s_1$ and
$s_2$ are the spins of the two incoming particles, allows us to
compare the $(\gamma N)$ channel, in a consistent way, to other
inelastic channels.

For the polarized photoproduction cross section with helicity $h$ we
have
\begin{equation}
\sigma_{\gamma,\pi}^h=\frac{2\pi}{k^2}(2J+1)\,|T_{\gamma,\pi}^h|^2,
\end{equation}
leading to the relation between unitary and helicity amplitudes
\begin{equation}
T_{\gamma,\pi}^h={\sqrt{2kq}}\; \mathcal{A}^{h}_\alpha\; C\,.
\label{uniamp}
\end{equation}

For a better understanding of the difference between Breit-Wigner
parameters and pole parameters, the unitary amplitude can be written
in terms of a propagator and initial and final partial
widths~\cite{pdg_kin},
\begin{equation}
T_{\gamma,\pi}^h(W)=\frac{(\Gamma_h/2)^{1/2}\;(\Gamma_\pi/2)^{1/2}}{M-W-i\Gamma/2} .
\label{NRansatz}
\end{equation}
At the Breit-Wigner resonance
position $W_r=M$ the amplitude becomes purely imaginary and the BW
resonance amplitude is defined as
\begin{eqnarray}
\tilde{T}_{\gamma,\pi}^h &=& \mbox{Im}\; T_{\gamma,\pi}^h(W_r)\\
&=& \frac{\Gamma_h^{1/2}\;\Gamma_\pi^{1/2}}{\Gamma}\,.\label{BWamp}
\end{eqnarray}

At the pole position, $W_p=M-i\Gamma/2$, the amplitude becomes
infinite and the pole parameter is defined as the complex residue
\begin{eqnarray}
R_{\gamma,\pi}^h &=& \mbox{Res}\; T_{\gamma,\pi}^h(W_p)\\
&=& \frac{\Gamma_h^{1/2}\;\Gamma_\pi^{1/2}}{2}\,.\label{Polamp}
\end{eqnarray}
Note that, traditionally, the residues of baryon resonances have
been defined with a relative minus sign compared to the standard
mathematical definition.

The complex residue can be factorized in
\begin{equation}
R_{\gamma,\pi}^h = \sqrt{\mbox{Res}\; T_{\gamma N}^h(W_p)\;
\mbox{Res}\; T_{\pi N}(W_p)}\,. \label{resfact}
\end{equation}
For these residues we will use in the following the short hand
notation $Res_{\gamma(h)N}$ and $Res_{\pi N}$.

At the BW position, the total photoproduction cross section with
helicity $h$ is
\begin{equation}
\sigma_{\gamma,\pi}^h=\frac{2\pi}{k^2}(2J+1)\,\frac{\Gamma_h\;\Gamma_\pi}{\Gamma^2}
\end{equation}
and the unpolarized cross section is
\begin{equation}
\sigma_{\gamma,\pi}=\frac{\pi}{k^2}(2J+1)\,(\Gamma_{1/2}+\Gamma_{3/2})\,\frac{\Gamma_\pi}{\Gamma^2}\,.
\end{equation}
With the relation~\cite{cko,pdg_res} between the $e.m.$ width and
the photo-decay amplitudes $A_{1/2}, A_{3/2}$
\begin{equation}
\Gamma_\gamma(M_r)=\frac{k^2}{\pi}\frac{2}{2J+1}\frac{m_N}{M_r}(|A_{1/2}|^2+|A_{3/2}|^2)\label{Gammagamma}
\end{equation}
the total cross section takes the form
\begin{equation}
\sigma_{\gamma,\pi}(M_r)=\frac{2m_N\Gamma_{\pi ,r} }{M_r
\Gamma_r^2}(|A_{1/2}|^2+|A_{3/2}|^2)\, ,
\end{equation}
where $\Gamma_r$ and $\Gamma_{\pi , r}$ are widths evaluated at the BW resonance
energy $M_r$ and $m_N$ is the nucleon mass.

Eq.~(\ref{Gammagamma}) can be used as a definition for the
photo-decay amplitudes
\begin{equation}
A_h = \sqrt{\frac{\pi(2J+1)M}{2k^2m_N}}\;\Gamma_h^{1/2}
\end{equation}
and by comparison with Eqs.~(\ref{uniamp},\ref{BWamp}) we obtain the
amplitudes at the Breit-Wigner position
\begin{equation}
A_h^{BW}=C\,\sqrt{\frac{q_r}{k_r}\frac{\pi(2J+1)M_r\Gamma_r^2}{m_N\Gamma_{\pi,r}}}\;
\tilde{\mathcal{A}}_\alpha^h\,.
\end{equation}
Similarly, a comparison with
Eqs.~(\ref{uniamp},\ref{Polamp},\ref{resfact}) leads to the
amplitudes at the pole position
\begin{equation}
A_h^{pole}=C\,\sqrt{\frac{q_p}{k_p}\frac{2\pi(2J+1)M_p}{m_N Res_{\pi
N}}}\; \mbox{Res}\,\mathcal{A}_\alpha^h\, ,
\end{equation}
where the subscript $p$ denotes quantities evaluated at the pole
position. 
The pole mass $M_p$ is the real part of the pole 
position $W_p$~\cite{M_p}.

Finally, normalized residues, partial widths, and branching ratios at
the pole can also be determined in accordance to the conventions of
the PDG.

The normalized residues are the residues divided by the half-width
at the pole,
\begin{equation}
(NR)_{\gamma,\pi}^h=\frac{R_{\gamma,\pi}^h}{\Gamma_p/2}=
\frac{\Gamma_h^{1/2}\;\Gamma_\pi^{1/2}}{\Gamma_p}\,,
\end{equation}
and obtain complex values, whereas the partial widths of the $\pi N$
and $\gamma N$ channels,
\begin{eqnarray}
\Gamma_{\pi,p} &=& 2 |Res_{\pi N}|\,,\\
\Gamma_{h,p}  &=& 2
|Res_{\gamma(h)N}|=\frac{2\,|R_{\gamma,\pi}^h|^2}{|Res_{\pi N}|}\,,
\end{eqnarray}
and the branching ratios at the pole
\begin{eqnarray}
BR_{pole}(\pi N) &=&
\frac{\Gamma_{\pi,p}}{\Gamma_p}=\frac{|Res_{\pi N}|}{\Gamma_p/2}\,,\\
BR_{pole}(\gamma^h N) &=&
\frac{\Gamma_{h,p}}{\Gamma_p}=\frac{|R_{\gamma,\pi}^h|^2}{|Res_{\pi
N}|\,\Gamma_p/2}\,
\end{eqnarray}
acquire real and positive numerical values.
\section{A simple model test}
\label{sec:said90}

In the first SAID analysis of pion photoproduction
data~\cite{vpi90},
multipoles were fitted using the form
\begin{equation}
M_{\ell} = \left( Born + B(W) \right) \left( 1 + i
T_{\pi N}^{\ell} \right) + C(W) T_{\pi N}^{\ell} ,
\label{oldform}
\end{equation}
based on a simple K-matrix approach~\cite{km}. This form had the
advantage that only the elastic $\pi N$ T-matrix was required (
$T_{\pi N}^{\ell}$ ), as connected to photoproduction via Watson's
theorem below the $\pi \pi N$ threshold, and continuing smoothly
from this constraint as the $\pi N$ partial waves became inelastic.
In the above, $\ell$ is the relative $\pi N$ angular momentum.
Labels for isospin and total spin have been suppressed. The
phenomenological pieces, $B(W)$ and $C(W)$ were polynomials in
energy with the required threshold behavior, and were fitted for
each partial wave.

In deriving Eq.~(\ref{oldform}), the inelasticity was assumed to be
dominated by a single channel. This simple approach has now been
improved~\cite{cm}. However, given a known set of elastic residues
and pole positions for the underlying $\pi N$ reaction, the above
form provides a simple test case for extracting pole-related
quantities in pion photoproduction, while giving a reasonable fit to
data. This fit has been reproduced for the present study.

The $\pi N$ T-matrix terms in Eq.~(\ref{oldform}) contain
information regarding included resonances and opening
thresholds~\cite{sp06,wi08}. As a result, the energy dependent
pre-factors are quite smoothly varying and can be represented by
low-order polynomials in energy. Here, and in Ref.~\cite{vpi90}, the
multipole residues were extracted from the known $\pi N$ pole
positions and residues, and a straightforward evaluation of the
energy-dependent pre-factors at the pole position.

Beyond being just a toy model, the form in Eq.~(\ref{oldform}) was
fitted to data from the $\pi^+ n$ threshold to a lab photon energy
of 2 GeV, sufficient to compare with the results of
Ref.~\cite{vpi90} and other more recent determinations~\cite{bnga}.
Results for both Breit-Wigner plus background and pole
determinations are given in Table I. The form of
background-resonance separation is very similar to that used in the
MAID fits~\cite{maid2007}, and is detailed in Ref.~\cite{bwfit}.
Errors for the Breit-Wigner fits were determined by fitting the
multipoles from the form of Eq.~(\ref{oldform}), 
using a Breit-Wigner resonance,
over varying energy ranges. For the pole determinations, the Born +
$B(W)$ and $C(W)$ were represented by two polynomials, $\alpha (W)$
and $\beta (W)$, of varying orders, over a range of energies
sufficient for extrapolation to the pole. Stability of these
results, and errors from the $\pi N$ elastic pole determinations,
were combined in a representative error.

\begin{table*}[ht]
\begin{tabular*}{\textwidth}{@{\extracolsep{\fill}}c|cccc|cccc}
\hline
 & \multicolumn{4}{c}{Breit-Wigner Values} & \multicolumn{4}{c}{Pole Values} \\
\hline
Resonance & (Mass,Width) & $\Gamma_\pi/2$ & $A_{1/2}$ & $A_{3/2}$ &
 (Re$W_p$ , $-2$Im $W_p$) & $R_\pi$ & $A_{1/2}$ & $A_{3/2}$\\
\hline
$\Delta(1232)\;3/2^+$ & $(1233 \; , \; 119)$ & $60$ & $-141\pm 3$ & $-258\pm 5$
         & $(1211 \; , \; 99)$ & $52\;[-47^{\circ}]$& $-135\pm 5\;[-17^{\circ}] $ & $-255\pm 5\;[-5^{\circ}]$ \\
$N(1440)\;1/2^+$ & $(1485 \; , \; 284)$  & $112$ & $-60\pm 2$ &
         & $(1359 \; , \; 162)$ & $38\;[-98^{\circ}]$ & $-65\pm 5\;[-37^{\circ}] $ &  \\
$N(1520)\;3/2^-$ & $(1515 \; , \; 104)$  & $33$ & $-19\pm 2$ & $+153\pm 3$ &
          $(1515 \; , \; 113)$ & $38\;[-5^{\circ}]$ & $-22\pm 3\;[-10^{\circ}] $ & $+156\pm 6\;[+11^{\circ}] $   \\
$N(1535)\;1/2^-$ & $(1547 \; , \; 188)$  & $34$  & $+92\pm 5$ &
         & $(1502 \; , \; 95)$ & $16\;[-16^{\circ}]$ & $+77\pm 5\;[+5^{\circ}] $ &  \\
$N(1650)\;1/2^-$ & $(1635 \; , \; 115)$ & $58$  & $+35\pm 5$ &
         & $(1648 \; , \; 80)$  & $14\;[-69^{\circ}]$ & $+36\pm 3\;[-16^{\circ}] $ &  \\
\hline
\end{tabular*}
\caption{\label{tab:pole1} Breit-Wigner and pole values for selected nucleon resonances.
Masses, widths and residues are given in
units of $MeV$, the helicity 1/2 and 3/2 photo-decay amplitudes in units of
$10^{-3} \times (GeV)^{-1/2}$. Errors on the phases are generally $2-5$ degrees.
For isospin 1/2 resonances the values of the proton target are given.}
\end{table*}

\section{Results and Conclusions}
\label{sec:results}

From Table I, we see that the pole and Breit-Wigner determinations,
for the states considered in Ref.~\cite{vpi90} plus a nearby state,
are quite similar in modulus. In the earlier determination, however,
the pole 'widths', constructed from squares of the helicity
amplitudes were found to be qualitatively similar for the $\Delta
(1232)$ and $N(1520)$, but radically different for the $N(1440)$ and
$N(1535)$ - differing in the latter cases by factors of about 2 and
5 respectively. A possible cause of the discrepancy is seen in the
$N(1440)$ and $N(1535)$ pole positions and elastic residues. From
1990 to present, the modulus of the elastic residue has shifted from
108 to 38 MeV, for $N(1440)$, and from 54 to 16 MeV, for the
$N(1535)$. For these states, the pole positions have also shifted
significantly.

For the pole amplitudes extracted in Table I, results are generally
quite similar to those from the Bonn-Gatchina group~\cite{bnga}. An
exception is the $N(1535)$, where the simple model of
Ref.~\cite{vpi90} is known to differ significantly from a more
sophisticated approach~\cite{bnga,cm}. A substantial benefit from
the pole extraction is found for the nearby $N(1650)$. BW fits to
the underlying $\pi N$ amplitude have produced unreliable
width/elasticity values which, in turn, have made BW fits to
photoproduction multipoles difficult and similarly unreliable. None
of these issues affect the pole determination.

Prior to the recent work of the Bonn-Gatchina group~\cite{bnga},
significant differences were seen in comparing photoproduction
amplitudes, determined through BW and pole determinations. These
included the early fits of Ref.~\cite{vpi90} as well as
determinations of the E2/M1 ratio~\cite{Han98,wor99} for the $\Delta
(1232)$. In the latter case, stability of this ratio at the pole was
found to be better than associated BW fits. Here we have repeated
the study of Ref.~\cite{vpi90}, finding results in qualitative if
not quantitative agreement with those of Ref.~\cite{bnga}. Results
are often quite similar to BW determinations, apart from a phase. In
cases where these values differ, the pole determination is more
reliable.

\begin{acknowledgments}
This work was supported in part by the U.S. Department of Energy
Grant DE-FG02-99ER41110, the Deutsche Forschungsgemeinschaft (SFB 1044),
and the RFBR grant 13-02-00425.
\end{acknowledgments}
\eject

\begin{appendix}
\section{Examples}
\label{sec:examples} The following examples will illustrate the
derivation of the photo-decay amplitudes at the pole position.

\subsection{$\Delta(1232)\,3/2^+$}
For the $\Delta(1232)$ resonance, we obtain a pole position of
$W_p=(1.211-i 0.099/2) \mbox{GeV}$ with an elastic residue of
$Res_{\pi N}=52\;e^{-i 47^\circ} \mbox{MeV}$. For photoproduction,
there are two isospin $3/2$ multipoles, for which we find the
residues
\begin{eqnarray}
\mbox{Res}\;M_{1+}^{3/2}&=&2.96\;e^{-i 30^\circ}\; \mbox{mfm}\,\mbox{GeV}\,,\\
\mbox{Res}\;E_{1+}^{3/2}&=&-0.16\;e^{i 35^\circ}\; \mbox{mfm}\,
\mbox{GeV}\,.
\end{eqnarray}
With Eqs.~(\ref{helimult1},\ref{helimult2}) we obtain the residues of the helicity
multipoles
\begin{eqnarray}
\mbox{Res}\;\mathcal{A}_{1+}^{1/2}
&=& -\frac{1}{2}(\mbox{Res}M_{1+}^{3/2} + 3\,\mbox{Res}E_{1+}^{3/2})\\
&=&-1.40\;e^{-i 39^\circ}\; \mbox{mfm}\,\mbox{GeV}\,,\\
\mbox{Res}\;\mathcal{A}_{1+}^{3/2}
&=& -\frac{\sqrt{3}}{2}(\mbox{Res}M_{1+}^{3/2} - \mbox{Res}E_{1+}^{3/2})\\
&=&-2.63\;e^{-i 27^\circ}\; \mbox{mfm}\,\mbox{GeV}\,.
\end{eqnarray}
In order to obtain the photo-decay amplitudes, these residues must
be multiplied by a complex factor depending on spin, isospin,
kinematics at the pole and the elastic $\pi N$ residue,
\begin{eqnarray}
A_h^{pole}&=&N\; \frac{\mbox{Res}\;\mathcal{A}_{1+}^{h}}{197\,
\mbox{mfm}\,\mbox{GeV}}\,,\\
N&=&C\sqrt{\frac{q_p}{k_p}\frac{2\pi(2J+1)M_p}{m_N \mbox{Res}_{\pi
N}}}
\end{eqnarray}
With the isospin factor $C=\sqrt{2/3}$, $q_p/k_p=0.88\,e^{-i
3^\circ}$, $J=3/2$, the pole mass $M_p=1.211\,\mbox{GeV}$ (real part of the pole position),
the nucleon mass $m_N$, giving $N=19.2\,e^{i 22^\circ}\,\mbox{GeV}^{-1/2}$ we obtain
the photo-decay amplitudes at the pole
\begin{eqnarray}
A_{1/2}^{pole}&=&-0.136\,e^{-i17^\circ}\,\mbox{GeV}^{-1/2}\,,\\
A_{3/2}^{pole}&=&-0.255\,e^{-i5^\circ}\,\mbox{GeV}^{-1/2}\,.
\end{eqnarray}
The magnitudes are very close to the Breit-Wigner values,
the phases are considerably smaller than the phases of the residues
themselves, because a large phase of the elastic residue is already
taken out.

\subsection{$N(1440)\,1/2^+$}
For the Roper resonance $N(1440)$, we obtain a pole position of
$W_p=(1.359-i 0.162/2)\, \mbox{GeV}$ with an elastic residue of
$Res_{\pi N}=38\,e^{-i 98^\circ}\, \mbox{MeV}$. For photoproduction,
there is only one isospin $1/2$ multipole, for which we find the
residue
\begin{eqnarray}
\mbox{Res}\;M_{1-}^{1/2}&=&0.35\,e^{-i 85^\circ}\; \mbox{mfm}\,\mbox{GeV}\,.
\end{eqnarray}
With Eq.~(\ref{helimult3}) we obtain the residue of the helicity
$1/2$ multipole as
\begin{eqnarray}
\mbox{Res}\;\mathcal{A}_{1-}^{1/2} &=& \mbox{Res}M_{1-}^{1/2}\,.
\end{eqnarray}
With the isospin factor $C=-\sqrt{3}$ and $q_p/k_p=0.95\,e^{-i
1^\circ}$, $J=1/2$, the pole mass $M_p=1.359\,\mbox{GeV}$,
giving $N=-37\,e^{i 48^\circ}\,\mbox{GeV}^{-1/2}$ we obtain
the photo-decay amplitude at the pole
\begin{eqnarray}
A_{1/2}^{pole}&=&-0.066\,e^{-i37^\circ}\,\mbox{GeV}^{-1/2}\,,
\end{eqnarray}
again a value with a magnitude close to the BW value and a much
smaller phase compared to the multipole residue.

\subsection{$N(1520)\,3/2^-$}
For the $D_{13}$ resonance $N(1520)$, we obtain a pole position of
$W_p=(1.515-i 0.113/2)\,\mbox{GeV}$ with an elastic residue of
$Res_{\pi N}=38\,e^{-i 5^\circ}\,\mbox{MeV}$. For photoproduction,
there are two isospin $1/2$ multipoles, for which we find the
residues
\begin{eqnarray}
\mbox{Res}\;E_{2-}^{1/2}&=&0.442\,e^{i 10.5^\circ}\; \mbox{mfm}\,\mbox{GeV}\,,\\
\mbox{Res}\;M_{2-}^{1/2}&=&0.196\,e^{i 4.5^\circ}\; \mbox{mfm}\,\mbox{GeV}\,.
\end{eqnarray}
With Eqs.~(\ref{helimult3},\ref{helimult4}) we obtain the residues of the helicity
$1/2$ and $3/2$ multipoles as
\begin{eqnarray}
\mbox{Res}\;\mathcal{A}_{2-}^{1/2}
&=& -\frac{1}{2}(\mbox{Res}E_{2-}^{1/2} - 3\,\mbox{Res}M_{2-}^{1/2})\\
&=&0.078\;e^{-i 13^\circ}\; \mbox{mfm}\,\mbox{GeV}\,,\\
\mbox{Res}\;\mathcal{A}_{2-}^{3/2}
&=& -\frac{\sqrt{3}}{2}(\mbox{Res}E_{2-}^{1/2} + \mbox{Res}M_{2-}^{1/2})\\
&=&-0.55\;e^{i 9^\circ}\; \mbox{mfm}\,\mbox{GeV}\,.
\end{eqnarray}
With the isospin factor $C=-\sqrt{3}$, $q_p/k_p=0.97\,e^{-i
0^\circ}$, giving $N=-56\,e^{i 2^\circ}\,\mbox{GeV}^{-1/2}$ we obtain
the photo-decay amplitudes at the pole
\begin{eqnarray}
A_{1/2}^{pole}&=&-0.022\,e^{-i 10^\circ}\,\mbox{GeV}^{-1/2}\,,\\
A_{3/2}^{pole}&=&+0.156\,e^{i 11^\circ}\,\mbox{GeV}^{-1/2}\,,
\end{eqnarray}
also values with magnitudes very similar to the BW values.

\subsection{$N(1535)\,1/2^-$}
For the $S_{11}$ resonance $N(1535)$, we obtain a pole position of
$W_p=(1.502-i 0.095/2)\,\mbox{GeV}$ with an elastic residue of
$Res_{\pi N}=16\,e^{-i 16^\circ}\,\mbox{MeV}$. For photoproduction,
there is only one isospin $1/2$ multipoles, for which we find the
residue
\begin{eqnarray}
\mbox{Res}\;E_{0+}^{1/2}&=&0.25\,e^{-i 3^\circ}\; \mbox{mfm}\,\mbox{GeV}\,,
\end{eqnarray}
With Eq.~(\ref{helimult1}) we obtain the residue of the helicity
$1/2$ multipole as
\begin{eqnarray}
\mbox{Res}\;\mathcal{A}_{0+}^{1/2} &=& -\mbox{Res}E_{0+}^{1/2}
\end{eqnarray}
With the isospin factor $C=-\sqrt{3}$, $q_p/k_p=0.97\,e^{-i
0^\circ}$, giving $N=-60\,e^{i 8^\circ}\,\mbox{GeV}^{-1/2}$ we obtain
the photo-decay amplitude at the pole
\begin{eqnarray}
A_{1/2}^{pole}&=&0.077\,e^{i5^\circ}\,\mbox{GeV}^{-1/2}\,,
\end{eqnarray}
a value with a magnitude which differs about $20\%$ from the BW
value.

\end{appendix}


\eject

\begin{thebibliography}{99}
\bibitem{M_p}
In Refs.\cite{vpi90} and \cite{bnga}, $M_p$ was replaced
by $W_p$.
\end{thebibliography}
\end{document}